# Unusual phase formation in reactively sputtered La-Co-O perovskite thin films

Tobias H. Piotrowiak[1], Rico Zehl[1], Ellen Suhr[1,] Lars Banko[1], Benedikt Kohnen[1], Detlef Rogalla[2], Alfred Ludwig[1]

[1]Chair for Materials Discovery and Interfaces, Institute for Materials, Ruhr University Bochum, Universitätsstr. 150, 44780 Bochum, Germany

[2] RUBION, Ruhr University Bochum, Universitätsstr. 150, 44780 Bochum, Germany

Corresponding author: alfred.ludwig@rub.de

## Abstract

La-based perovskites are a versatile class of materials that are of interest for solid oxide fuel cells and electrocatalytic water splitting. During fabrication of composition spread materials libraries of La-Co-based oxide systems for the discovery of new catalytic materials, an unusual phase formation phenomenon was observed: instead of the expected continuous composition gradient, regions with homogeneous composition and single-phase structure ($La_2O_3$ or stoichiometric La-perovskite) form. This phenomenon occurs during reactive co-sputtering and is dependent on $O_2$-flux and substrate temperature, investigated from room temperature up to 700 °C and is independent of the used substrate. It can be described as a self-organized growth, where excess transition metal cannot be incorporated into the growing film and the forming single-phase regions. It is hypothesized that due to the high reactivity of La and the significantly low formation energies of $La_2O_3$ and La-perovskites, the reactive sputter deposition of La-based oxide films can turn – regarding film growth – into a partial CVD-like process which results in the unusual self-organized growth of single-phase regions. This phenomenon can be leveraged for the exploration of multinary perovskite thin film libraries, where the B-site atoms of La-perovskites are systematically substituted.

# Introduction

Reactive magnetron sputtering is well-established for the fabrication of multinary compound thin films. In dependence of the used targets and sputter parameters, oxide films with defined compositions and phases can be prepared. Combinatorial magnetron sputtering using co-deposition from multiple targets allows the fabrication of materials libraries (MLs) which seamlessly cover large compositional spaces that are subsequently screened for their properties [1–3]. This enables efficient investigations of large compositional spaces. Subsequently, hundreds of compositions on a materials library are automatically characterized for their structural and functional properties using suitable high-throughput methods, as presented in the experimental section of this work. The compositions of a fabricated ML changes continuously along the gradient direction with typical lateral gradients of 0.3 - 1 (at.% / mm) [4–8]. Co-sputtered binary composition spreads usually show a continuous, often quasi-linear change in chemical composition along the gradient, independent of its metal- [9–11], oxide- [5,12–14] or nitride-state [15,16].

La-based perovskites are promising materials that are of interest for many applications, e.g. in solid oxide fuel cells (SOFC) [17,18] and electrocatalytic water splitting [19–21]. By doping the perovskite, e.g. with Sr, electrical conductivity can be improved [19]. By tuning the chemical composition of La-based perovskites, the catalytic activity for the oxygen evolution reaction (OER) and oxygen reduction reaction (ORR) can be improved [21]. In the context of this research, compositional complexity of the perovskites is increasing, up to high-entropy perovskite materials [22–24]. Although La-Co-O is a well explored system [25], the influence of reactive magnetron sputtering parameters on composition, structure and properties have not been studied systematically over large compositional ranges. Moreover, only a few reports on combinatorial approaches for the sputter fabrication and characterization of La-based perovskite forming systems like $La_xTa_{1-x}O_yN_z$ [26], $LaFeO_3$ and $LaFeO_{3-x}N_x$ [27] exist. Most studies focused on fabrication of discrete perovskite compositions or sputtering from a perovskite-composition target, like reported for $LaCoO_3$ [28,29], La-Cu-O [30], La-Ni-O [31,32], $LaMnO_3$ [33], La-Pb-Mn-O [34], $LaVO_3$ [35], La-Ti-O [36] or $LaCrO_3$ [37]. Since none of the studies investigated a larger compositional range of the described La-M-O (M = transition metal) systems, no unusual film growth or phase formation behavior such as reported in this work, was observed. As the combinatorial approach is well-suited to fabricate large compositional spaces and tailor material properties, we use a co-deposition sputter system to study the influence of parameter variations, i.e. temperature, sputter power, $O_2$-flux or process pressure, on properties of La-Co-O thin film MLs. Hereby, we discovered a $O_2$-flux and temperature dependent film formation phenomenon, where stoichiometric and single-phase perovskites grow in a self-organized manner on different inert substrates.

# Experimental methods

## Synthesis of La-Co-O thin film libraries

La-Co-O libraries were fabricated by reactive ($O_2$/Ar atmosphere) magnetron co-sputtering in a commercial sputter system (AJA International, ATC 2200 with four confocally aligned sputter sources). La was sputtered from a compound target, $La_2O_3$ (4-inch diameter, 99.99 % purity, Evochem Advanced Materials GmbH) and Co from an elemental Co-Target (4-inch diameter, 99.99 %, Sindlhauser Materials GmbH). For the fabrication of pseudo-binary La-Co-O composition spreads, two sputter cathodes were used, schematically shown in Figure 1a): to get composition gradients covering the maximal achievable composition range, the two cathodes directing to the substrate center, were positioned 180° from each other (Figure 1b)).

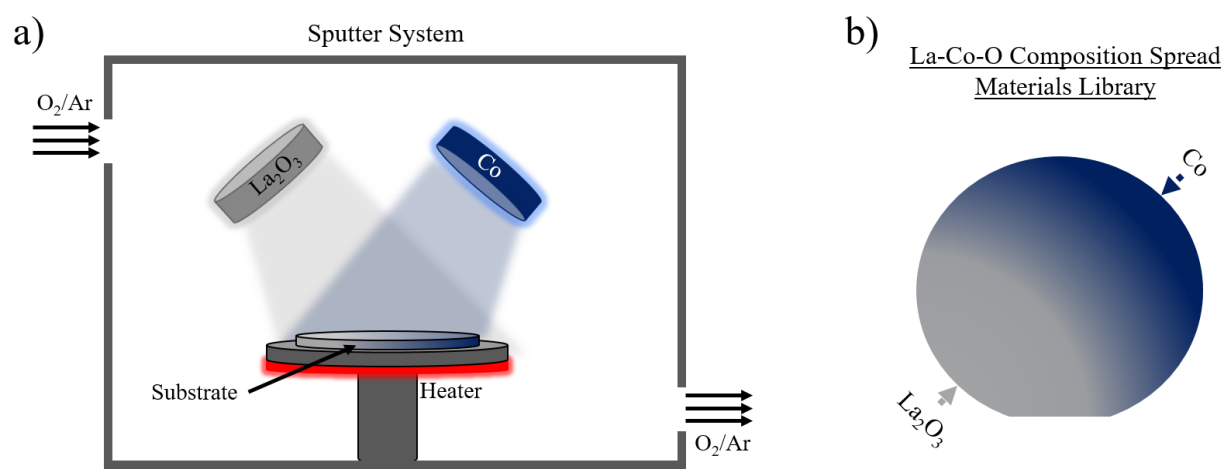

Figure 1: a) Schematic illustration of the reactive magnetron co-deposition process to fabricate materials libraries of oxide thin film systems. b) A schematic of the binary ML deposited on a 4-inch diameter wafer is shown and the sputter target positions are indicated.

Depending on the desired composition, $La_2O_3$ was sputtered with a RF power of 200 W for La-rich or 100 W for Co-rich libraries, while Co was always sputtered with a DC power 70 W. As substrates 4-inch diameter oxidized Si-wafers with 500 nm of $SiO_2$ (diffusion barrier) and sapphire or strips cut from these substrates were used. All depositions that showed the unusual phenomenon were carried out at a fixed process pressure of 0.4 Pa, with an $O_2$/Ar flow ratio of 40 sccm / 80 sccm. While investigating the influence of different process parameters, lower $O_2$-flows (< 40 sccm) were used as well. Substrate temperatures during depositions ranged from room temperature (i.e. no intentional heating) to 700 °C.

**High-throughput characterization**

For high-throughput characterization of the ML a fixed and uniform measurement scheme was used, to apply all methods to the same measurement areas (MAs) and make results comparable between each other: for all La-Co-O ML, measurements were started approximately 2 mm from the edge of the La-rich side and were performed in 4.5 mm steps towards the Co-rich side. This way a maximum of 22 MAs was realized on the 4-inch wafer or strips, cut from these substrates. The actual size of each investigated MA is dependent on the used measurement method.

Film thickness was measured using a tactile profilometer (Ambios XP2) on steps of partially removed film. These steps were created using photolithographic techniques or by partially covering the substrate during film deposition.

The chemical composition, with respect to the metals and omitting oxygen, was analyzed using automated electron-dispersive X-ray spectroscopy (EDX) in a scanning electron microscope (SEM, JEOL 5800 equipped with an Oxford INCA X-act detector). The SEM was operated at 20 kV with a magnification of 600, leading to a measured area of 400 μm x 600 μm. The composition of these MAs is averaged over the area and considered homogeneous within the MA. Measurement data was normalized on La and Co and is given in at. %. All compositions measured with EDX are described and discussed on the basis of the metallic gradient (La-content = La/(La+Co) or Co-content Co/(La+Co)), since the O-signal is biased by the oxide substrate and therefore cannot be quantified correctly.

For determination of O contents (O/(La+Co+O)) in the films elastic backscattering spectroscopy (EBS) was performed on selected samples at the 4 MV accelerator facility of RUBION (Ruhr University Bochum). The measurements were performed using a doubly charged 5.05 MeV He beam collimated to diameter of 1 mm and with intensities of about 50 nA. The samples were tilted at an angle of 7 degrees. A Si surface barrier detector with a solid angle of 1.9 msrad was placed at 160° with respect to the beam axis. The spectra were analyzed using the SIMNRA software [38]. The error in the data determination is about 3 % for the metal composition and about 7 % for the O-determination.

The crystallographic phase analysis was performed by X-ray diffraction (XRD). A Bruker D8 Discover with a Vantec-500 2D-detector in Bragg-Brentano geometry and an Incoatec High Brilliance Iμs Cu $K_\alpha$ X-ray source (0.154059 nm) was used. Three Frames were taken stepwise at every MA with an increment of θ/2θ 10°/20°, starting at 12.5°/25° and finishing at 32.5°/65°. In case of Si-substrate measurements were performed with a 2.5° offset on θ to avoid single crystal Si-substrate peaks. This way an angular 2θ range from approximately 10° to 85° was covered. For visualization, the angular 2θ range of the diffraction pattern was limited from 15° to 80°. For peak indication Pearson's Crystal Database was used.

The surface microstructure was examined by SEM (JEOL JSM-7200F), operated at acceleration voltages ranging from 5 kV to 15 kV, depending on the thin film's conductivity, and with a working distance of about 5 mm. AFM images were recorded using a Bruker FastScan scanning probe microscope in PeakForce Tapping mode with ScanAsyst, using a FastScan A probe (spring constant approximately 18 N/m).

## Results and Discussion

### Correlations between composition and thickness profiles, optical appearance and surface microstructure

For the reactive co-deposition of the La-Co-O ML a continuous composition gradient was expected and observed for all sputter processes with small $O_2$/Ar ratios or without additional oxygen. This composition gradient has, referring to the metals, a continuous profile, as shown in Figure 2a) for an exemplary ML deposited at 500 °C. The same continuous profile was observed for the other substrate temperatures used, i.e. room temperature (RT), 300 °C and 700 °C under small $O_2$/Ar ratios.

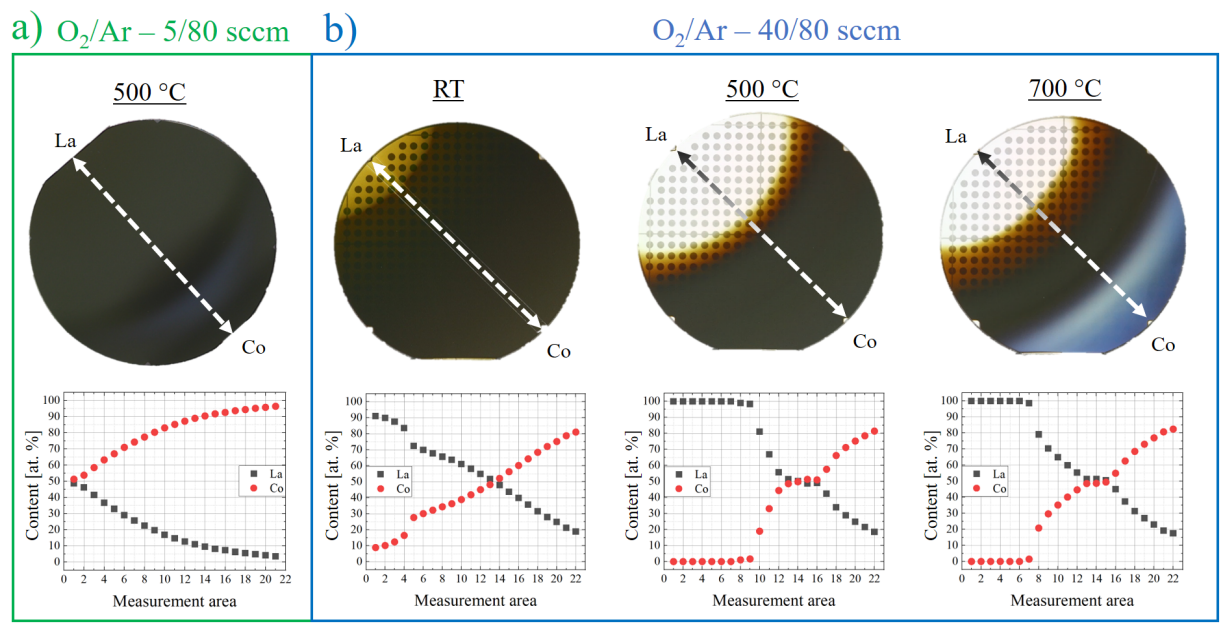

*Figure 2: Influence of oxygen-flow and substrate temperature on the formation of composition plateaus investigated on La-Co-O ML sputtered at temperatures ranging from room temperature to 700 °C at $O_2$/Ar ratios of a) 5 sccm / 80 sccm and b) 40 sccm / 80 sccm. Composition data is normalized on La and Co, omitting substrate and oxygen signals. The dots are from a pattern underneath the wafer, that is used to observe transparent regions in a ML.*

However, for reactive sputtering at high substrate temperatures and higher $O_2$/Ar ratios, where more oxygen for oxide phase formation is available, an unusual composition formation behavior was observed: instead of the expected gradient, large lateral regions with a surprisingly homogeneous composition formed, leading to discontinuous gradients, as shown in Figure 2b for libraries deposited at 500 °C and 700 °C. These regions have an arcuate shape on the wafer, that is following the La-particle flux from the La- to the Co-rich side. For depositions carried out at RT, i.e. without intentional heating, the composition profile shows an arcuate step on the La-rich side, while the rest of the gradient shows a quasi-linear profile. The La-rich area with a maximum La-content of 90 at. % appears yellowish-transparent, whereas the rest of the ML is opaque. With increasing substrate temperature, the composition profile changes significantly: The La-rich step in the profile expands and becomes a $LaO_x$, whitish-transparent area on the substrate. Moreover, at temperatures above 300 °C, a second region with homogeneous composition of about $(La_{50}Co_{50})O_x$ forms. At 500 °C these two plateaus reach their largest expansion. At 700 °C, the transition area between the $LaO_x$ plateau and the $(La_{50}Co_{50})O_x$ plateau enlarges, whereas the plateaus stay stable.

The non-normalized EDX data of the La-Co-O ML with discontinuous composition gradient shows that the Co content changes atypical along the measurement areas. Figure 3a) displays the metal content data of La-Co-O ML deposited at 500 °C: The La gradient is quasi-linear, whereas the Co content shows a discontinuous composition profile with pronounced non-linearities, which coincides with the plateaus of films with homogeneous composition in the normalized data. The results indicate that Co is not incorporated into the film as would have been expected. This leads to the assumption, that depending on the $O_2$-flux, surface adsorption and/or incorporation of Co in the growing film is changed.

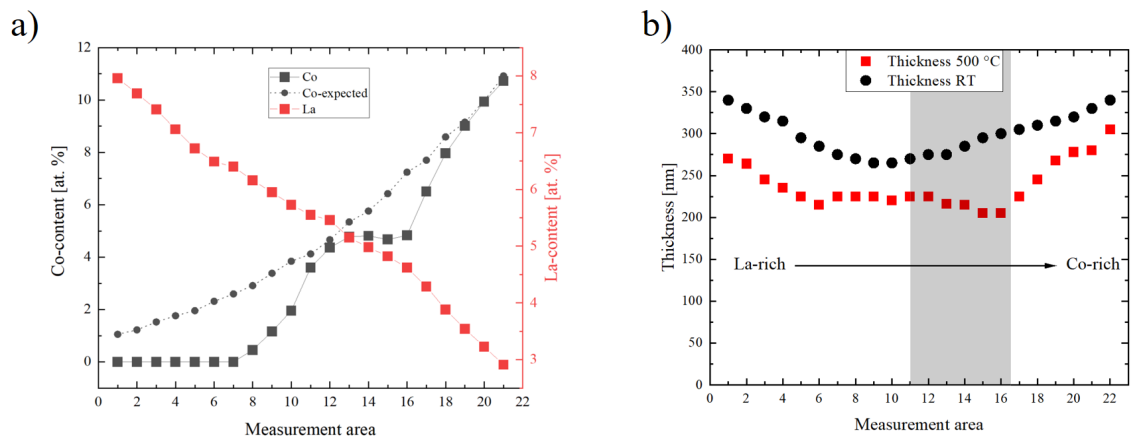

*Figure 3: a) Non-normalized La and Co compositional EDX raw data in at. %, measured on a La-Co-O ML deposited at 500 °C with O$_2$/Ar ratios 40 sccm / 80 sccm. La shows a quasi-linear composition gradient, whereas Co shows a step profile. The dotted black line shows the expected composition gradient of Co. b) Thickness gradients of La-Co-O ML fabricated at RT and 500 °C substrate temperature with O$_2$/Ar ratios of 40 sccm / 80 sccm. Thickness of the 500 °C ML decreases in the region with constant Co-content (highlighted in grey), whereas the thickness of the RT ML, that has a continuous Co-gradient there, increases.*

Film thickness values measured along the composition gradient resemble the profile observed in the EDX data for MLs deposited at higher temperatures. Figure 3b) compares the thickness profile of La-Co-O MLs deposited at RT and at 500 °C. The films of the ML fabricated at RT are generally about 10 – 30 % thicker than those sputtered at 500 °C. This can be partially explained by increasing desorption of sputtered material at elevated substrate temperatures. However, libraries deposited at different temperatures do not show the same thickness profile along their composition gradient. The greyish area in the thickness profile highlights the region, where $(La_{50}Co_{50})O_x$ forms at high temperatures. The thickness profile of the La-Co-O ML deposited at RT is quasi-symmetric: Thickness decreases continuously from the La-rich side to the library center and then increases continuously to the Co-rich side. The thickness of the La-Co-O library deposited at 500 °C has a comparable profile until reaching the plateau region, where the Co-content stays constant in the film: here, thickness stays constant or is even slightly decreasing. This can be rationalized, since the La-content in the composition gradient is decreasing and the Co-content stays approximately constant. When the Co-content increases again, also film thickness increases significantly.

In the temperature range from RT to 500 °C all La-Co-O MLs show a smooth and fine-grained surface microstructure. At 700 °C, the morphology of the single-phase LaCoO$_3$ region is still relatively smooth and fine-grained. However, for Co-contents > 50 at.%, large, separated grains/islands are observed as shown in Figure 4. These large structures agglomerate with increasing Co-content and develop into separated columns. The formation of these islands and columns leads to the bluish appearance of the ML shown in Figure 2b).

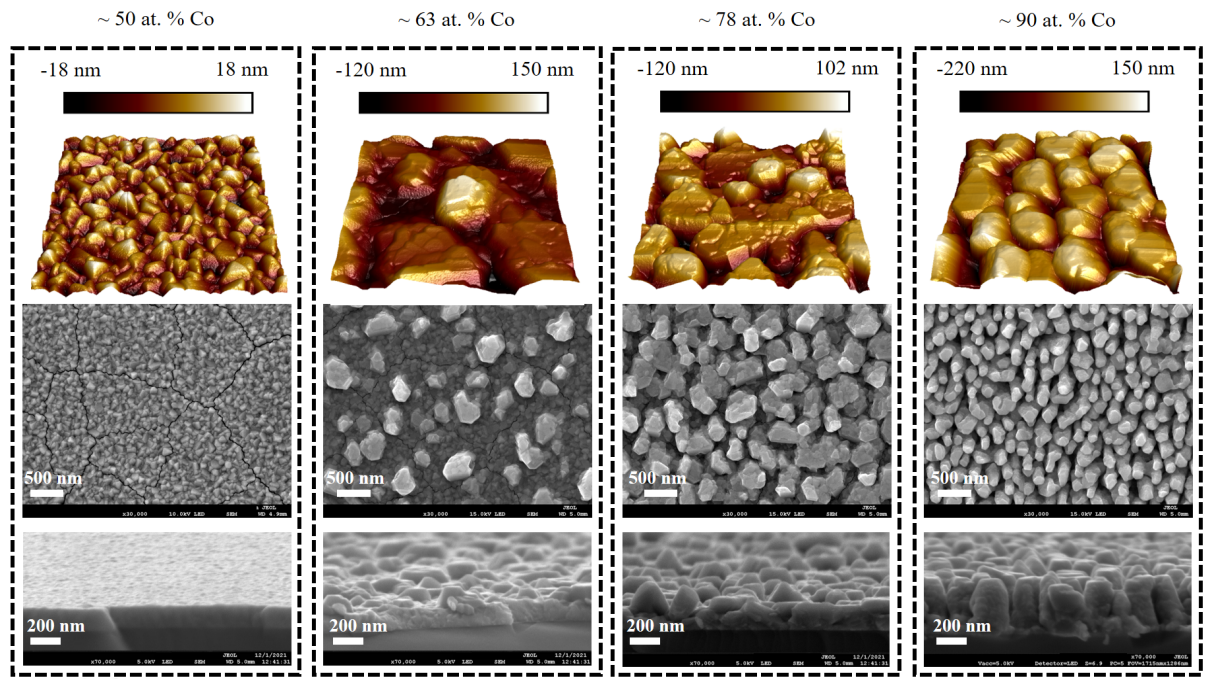

*Figure 4: AFM and SEM images from a La-Co-O ML fabricated at 700 °C with $O_2$/Ar ratios of 40 sccm / 80 sccm in the compositional range from 50 – 93 at.% Co on oxidized Si. Top view SEM images were taken with 30k magnification, cross-section SEM images were taken with 70k magnification and AFM images display a scanned area of 1 µm x 1 µm.*

## Explanation of composition- and phase-plateau formation - self-organized growth of $La_2O_3$ and $LaCoO_3$ films

XRD analysis of the La-Co-O ML reveals an unusual phase formation along the composition gradients. MLs deposited at RT are X-ray amorphous, except in the small La-rich yellowish-transparent areas of the ML with La-content up to 90 at. % (see Fig. 2b), where a weak XRD signal of $La_2O_3$ was detected. However, La-Co-O deposited at elevated substrate temperatures show single-phase formation in the regions where the homogeneous composition plateaus form. Figure 5 shows the XRD data as waterfall plot for an exemplary La-Co-O ML deposited at 500 °C. Three phase regions can be distinguished, depending on the chemical composition. $La_2O_3$ is detected in the pure La-O region (see Figure 5, highlighted in blue). With increasing Co-content, approaching $(La_{50}Co_{50})O_x$, the $La_2O_3$ signal vanishes and the XRD-signal of $LaCoO_3$ perovskite appears. Along this composition plateau (highlighted in grey) only the XRD-signal of $LaCoO_3$ perovskite is detected. With further increasing Co-content the $LaCoO_3$ signal decreases and a Co-oxide signal appears in the XRD patterns. For films deposited at 300 °C (not shown) and 500 °C, $Co_3O_4$ is detected, see Figure 5. For the ML fabricated at 700 °C a XRD-signal of CoO is detected (not shown), which is in agreement with known phase diagrams for La-Co-O [25]. The XRD results show that the regions with homogeneous composition (highlighted in blue and grey) are single phase and phase pure.

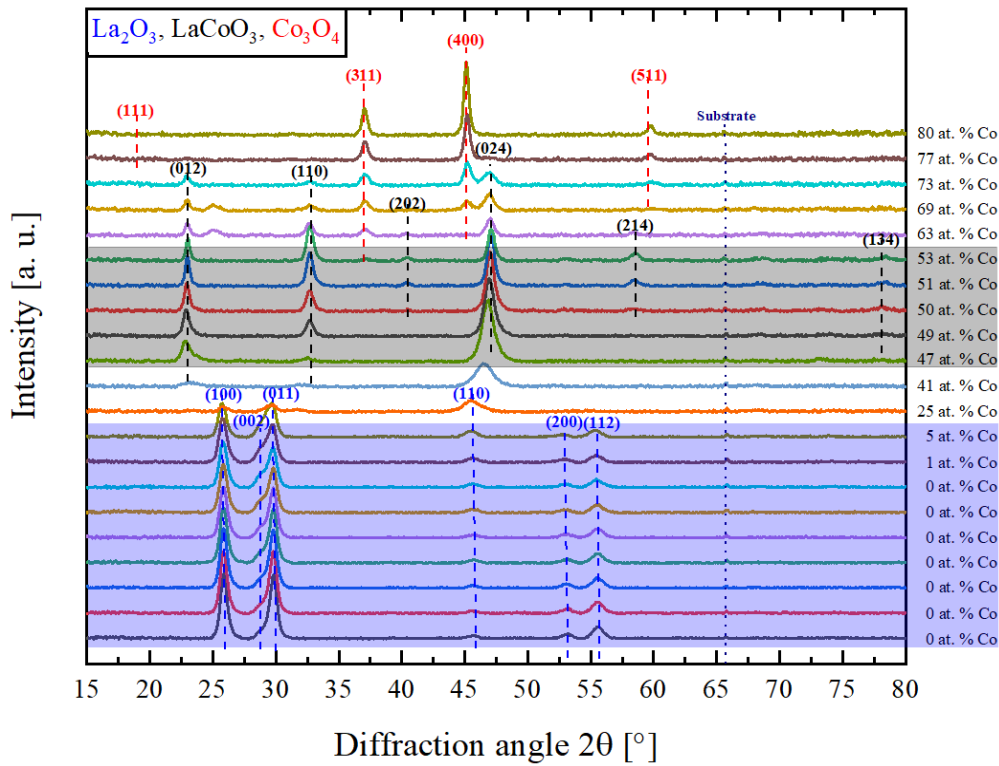

*Figure 5: Waterfall plot of XRD patterns along the composition gradient of a La-rich La-Co-O ML sputtered at an $O_2$/Ar ratio of 40 sccm / 80 sccm and 500 °C substrate temperature. The area highlighted in blue shows the regions of the $La_2O_3$-plateau, grey the area of the $LaCoO_3$ perovskite plateau. Indication of peaks using Pearson´s Crystal Data: $La_2O_3$ ID1221848, $LaCoO_3$ ID1920206, $Co_3O_4$ ID1824309*

Elastic backscattering spectroscopy (EBS) measurements of the film composition in the composition region with homogeneous $(La_{50}Co_{50})O_x$ plateau show an O-content of about 60 % for substrate temperatures from 300 °C to 700 °C. This complements the XRD-measurements and indicates that the $LaCoO_3$ perovskite forming in this area is stoichiometric.

In the sputter processes with low $O_2$/Ar ratios, i.e. 5 sccm / 80 sccm, where no regions with homogeneous composition plateau formed, only for the film composition of about $(La_{50}Co_{50})O_x$ pure $LaCoO_3$ signal is detected in XRD. With increasing Co-content in the continuous gradient (after the plateau) additional signal of Co-oxides appears, showing a multi-phase mixture along the composition gradient, see Figure 6.

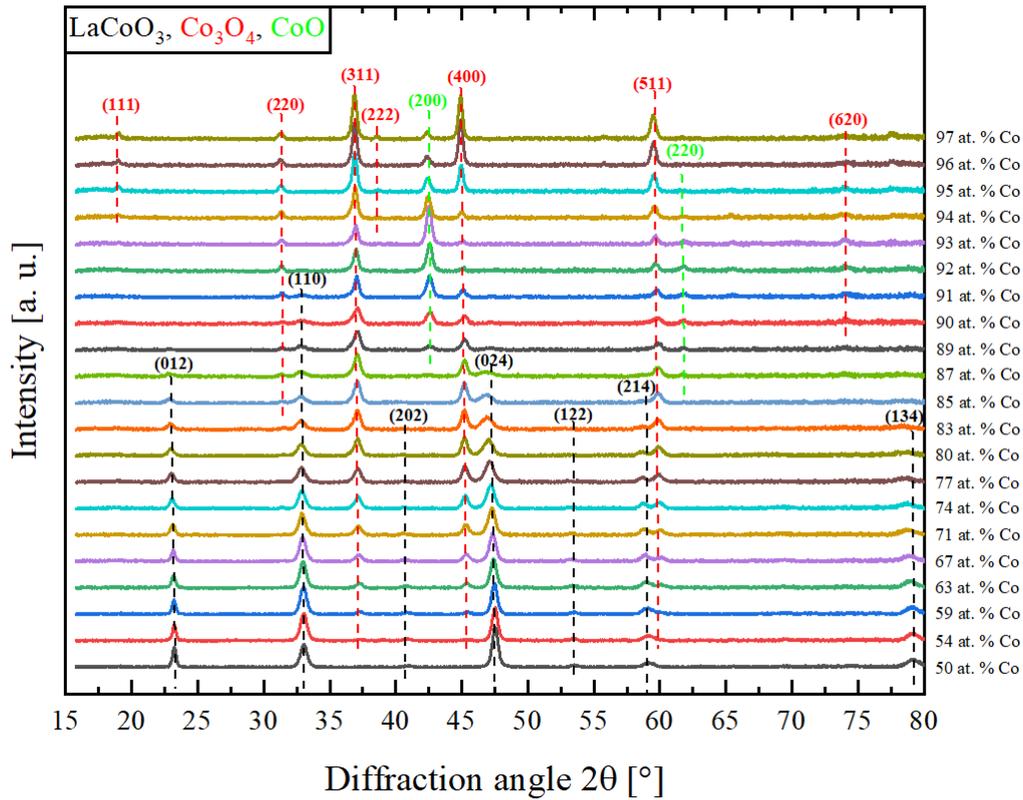

*Figure 6: Waterfall plot of XRD patterns along the composition gradient of a La-rich La-Co-O ML sputtered at an O$_2$/Ar ratio of 5 sccm / 80 sccm and 500 °C substrate temperature. Indication of peaks using Pearson´s Crystal Data: LaCoO$_3$ ID1920206, Co$_3$O$_4$ ID1824309, CoO ID1927666.*

A possible explanation for the formation of regions with homogeneous composition inside an intended composition gradient could be lateral surface diffusion. To verify or falsify this hypothesis, a series of experiments was performed at sputter conditions where the plateau formation was observed before. A La-Co-O ML was deposited at 500 °C on (I) a continuous "control substrate" and (II) on 5 mm wide substrate strips (cut from oxidized (100)-Si wafer) which were separated by a gap, see Figure 7. The distance between the strips was increased in several deposition experiments until a 5 mm wide gap between the strips was reached (iterations not shown). The working hypothesis for this experiment was that in case of the separated substrate strips (II) a possible lateral surface diffusion would be suppressed and thus a linear composition spread would be observed, whereas for continuous substrate (I) the steps and plateaus in the gradient should appear.

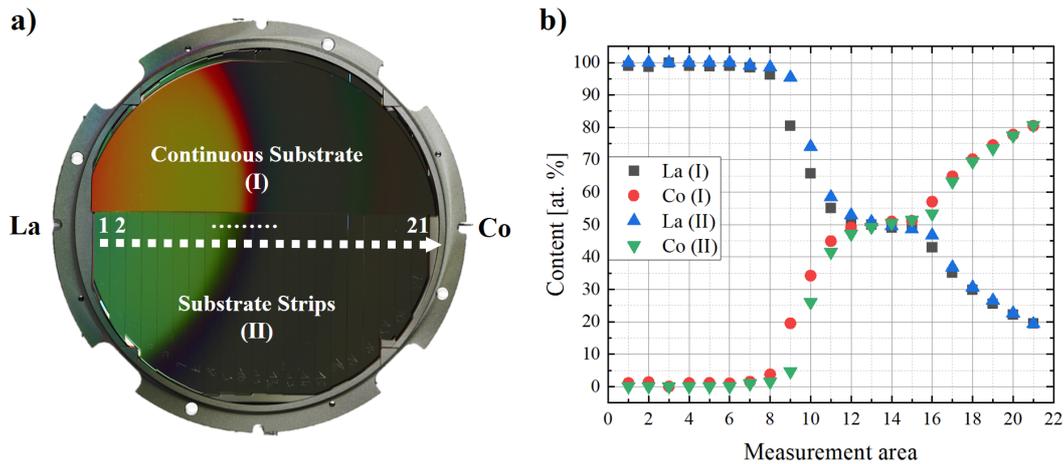

Figure 7: (a) Photograph of the La-Co-O ML deposited at 500 °C and O$_2$/Ar ratio of 40 sccm / 80 sccm on (I) the continuous control substrate and (II) separated oxidized Si substrate strips; (b) Results of EDX measurements of the La-Co-O ML on the (I) continuous control substrate and (II) the separated substrate strips. The dotted arrow indicates the measurement areas.

EDX analysis reveals the formation of the same compositional plateaus on the continuous reference substrate and over the separated substrate strip, <u>independent of the gap-width between the strips</u>. This proves, that the observed phenomenon cannot be caused by lateral surface diffusion and leads to the conclusion, that the composition-phase-formation behavior has to be gas phase related. Another hypothesis, which might be a suitable explanation for the observed phenomenon, is a kind of PVD-CVD hybridization of the deposition process. Hybridization hereby means, that the film material is delivered as particle flux by sputtering, whereas film growth and phase formation happen comparable to a CVD process with regard to reactivity of film forming materials and the temperature influence from the substrate: Following DFT calculations results retrieved from "The Materials Project" [39,40], La$_2$O$_3$ and LaCoO$_3$ are the most stable phases in the La-Co-O system with significantly low formation energies around -3.87 eV/atom and -2.6 eV/atom. Both phases are line compounds, that cannot incorporate additional amounts of elements beyond their stoichiometry. Therefore, they should coexist in a phase mixture (La$_2$O$_3$ + LaCoO$_3$) with mixing ratios according to the overall composition. [25] Pure La has a high affinity to electrons and oxygen [41], which favors the growth of these two phases in oxygen-rich atmospheres. Elevated substrate temperatures favor crystallization and phase formation as well. Moreover, during reactive sputtering in processes with high oxygen partial pressure and target poisoning, a large amount of high-energetic negative O$^-$-ions is generated in front of the target, that is accelerated in the electrical field to the substrate and the growing film [42–45]. The amount of O$^-$-ions generated during sputtering is dependent on the target condition, influenced by the oxygen partial pressure, as well as the target material itself. With increasing O-partial pressure, the amount of negative O$^-$-ions increases. These highly energetic, negative O$^-$-ions can have energies of several hundred eV up to the cathode voltage.

[43,45,46] In literature it is reported that the O$^-$-ion bombardment of the substrate and growing films, influences phase formation, stoichiometry and microstructure of the film. [42,47,48] The ion bombardment can even etch the substrate and growing film or lead to resputtering [47,49]. In this work, the used La$_2$O$_3$-target is in fully oxidized state (in sputter terminology "poisoned"), and therefore should generate a high amount of negative O$^-$-ions that get accelerated to the substrate and growing film. We assume that, due to the significantly low formation energies of La$_2$O$_3$ and LaCoO$_3$ and an abundant amount of O$^-$-ions being present on the substrate during film growth, these two phases are likely to form. The location of these phase regions forming on the wafer seems to be dependent on the amount of La-particles reaching the surface and the overall La/Co-particle ratio. For regions with La-rich particle flux reaching the substrate, pure La$_2$O$_3$ forms. For regions where the particle-flux ratio is in the range of (La$_{50}$Co$_{50}$)O$_x$, the LaCoO$_3$ perovskite forms. The dependency of the La-particle flux can also be seen in the arcuate shape of the single-phase region in sputter direction (see Figure 2). It can be further hypothesized, that excess Co adatoms, which are not incorporated in the film on lattice sites of the stoichiometric phases (e.g. B-site), have lower binding energy and desorb, or are preferentially resputtered. However, it could not be finally clarified, why the excess Co cannot be incorporated into the film and what exactly happens during growth. This phase formation is significantly affected by the amount of O$^-$-ions available in the plasma as well as at the substrate and further the substrate temperature. At low substrate temperatures there is not sufficient energy for a CVD-like film growth and phase formation. Similarly, at depositions with low O$_2$/Ar ratios there might be not sufficient O$^-$-ions available to create highly reactive conditions for a CVD like film growth and phase formation. This is clearly seen for the 500 °C deposition with low O-content, where a continuous composition spread forms and the self-organized phase formation is inhibited (see Figure 2a) and Figure 6. Summing the above, given the high amount of O$^-$-ions in the plasma and elevated substrate temperature results in a CVD-like, self-organized growth of the dominant, thermodynamically most stable phases La$_2$O$_3$ and LaCoO$_3$.

## Generality of the observed phenomenon in reactively sputtered La-based perovskite systems

Further experiments (not shown) with other La-transition metal combinations like Mn, Fe, Cr under identical deposition conditions show, that the same phenomenon occurs as observed for the La-Co-O system. For all three combinations the two homogeneous composition regions form with single-phase $La_2O_3$ and $LaMO_3$ (M = Mn, Fe, Cr) perovskite. Following DFT calculations from "The Materials Project" [39,40], also these perovskites have significantly low formation energies of about -2.99 eV/atom for $LaMnO_3$, -2.85 eV/atom for $LaFeO_3$, and -3.19 eV/atom for $LaCrO_3$. Based on this, a generality of the observed phenomenon for the fabrication of $LaMO_3$ perovskites under reactive sputter conditions can be concluded. Moreover, this self-organized growth can be leveraged for the combinatorial fabrication of perovskite thin films with continuous variation of perovskite composition, when sputtering multi-transition-metal-La-combinations, since the transitions metals (M) can substitute each other in the $LaMO_3$ perovskite structure. This could also be shown in first experiments with the La-Co-Mn-O and La-Co-Mn-Fe-O systems.

## Conclusions

During reactive co-sputtering of La-based transition metal oxide libraries, an unusual composition and phase formation behavior was observed. For processes with high $O_2/Ar$ ratios, formation of two large lateral regions with homogeneous composition and phase was reported along a composition gradient which was expected to be continuous. The formation is dependent on the substrate temperature and was investigated in the temperature range from RT to 700 °C. One region is La-pure single phase $La_2O_3$. The second region has the discrete metal ratio for perovskites of $La_{50}Co_{50}$ and stoichiometric single-phase $LaCoO_3$ forms. From lateral diffusion experiments, it could be excluded that the phenomenon is driven by surface diffusion. Variation of oxygen partial pressure during deposition is also revealing a strong oxygen dependency of the formation of these regions. From these observations it was concluded that the phenomenon is driven by a change from a PVD to a CVD-like film growth, due to the high reactivity of La, highly reactive $O^-$-ions in the process and the significantly low formation energies of $La_2O_3$ and $LaCoO_3$ in combination with the energy from heated substrates. The phases which form are dependent on the composition of the particle flux, that reaches the substrate. The phenomenon is interpreted as kind of self-organized growth in which excess transition metal (Co) that cannot be incorporated into the phase during film growth. It could not be finally clarified why the excess Co cannot be incorporated into the film. Moreover, a generality of the observed phenomenon can be concluded for La-M-O (M = transition metal) systems, since comparable self-organized growth of the both described single phase regions was observed for La-Mn-O, La-Fe-O, and La-Cr-O. The self-organized growth of La-perovskite can be leveraged for the combinatorial sputter-fabrication and investigation of La-based perovskites. Continuous single perovskite composition spreads can easily be fabricated and subsequently screened for their properties by sputtering multiple transition metals with La, since they can substitute each other in the $LaMO_3$ (M = transition metal) perovskite.

## Acknowledgements

This work was funded by the Deutsche Forschungsgemeinschaft (DFG, German Research Foundation) – Project number 388390466-TRR 247, project C04. Additionally, the SFB TR 87 and SFB TR 103 are acknowledged for support. ZGH at Ruhr-University Bochum is acknowledged for XRD, SEM and AFM measurements. RUBION at Ruhr-University Bochum is acknowledged for EBS measurements.